\documentclass[aps,pra,showpacs,twocolumn,amsmath,amssymb]{revtex4-1}
\usepackage{graphicx}
\usepackage{amsmath}
\usepackage{color}
\usepackage{dcolumn}
\usepackage{bm}
\usepackage{booktabs}
\usepackage{subfigure}

\newcommand{\fHe}{$^4$He$^*$}
\newcommand{\He}{He$^*$}
\newcommand{\etal}{\textit{et al.}}

\begin{document}

\title{Universal three-body parameter in ultracold $^4$He$^*$}

\author{S.\,Knoop,$^{1}$ J.\,S.\,Borbely,$^{1}$ W.\,Vassen,$^{1}$ S.\,J.\,J.\,M.\,F.\,Kokkelmans$^2$}

\affiliation{$^{1}$LaserLaB Vrije Universiteit, De Boelelaan 1081, 1081 HV Amsterdam, The Netherlands\\
$^{2}$Eindhoven University of Technology, P. O. Box 513, 5600 MB Eindhoven, The Netherlands}

\date{\today}

\begin{abstract}
We have analyzed our recently-measured three-body loss rate coefficient for a Bose-Einstein condensate of spin-polarized metastable triplet $^4$He atoms in terms of Efimov physics. The large value of the scattering length for these atoms, which provides access to the Efimov regime, arises from a nearby potential resonance. We find the loss coefficient to be consistent with the three-body parameter (3BP) found in alkali-metal experiments, where Feshbach resonances are used to tune the interaction. This provides new evidence for a universal 3BP, the first outside the group of alkali-metal elements. In addition, we give examples of other atomic systems without Feshbach resonances but with a large scattering length that would be interesting to analyze once precise measurements of three-body loss are available.
\end{abstract}


\maketitle

\section{Introduction}

When the short-range interaction between particles is very large, few-body properties are expected to become universal, i.\,e.\,, irrespective of the precise nature of the interaction and therefore applicable to nucleons, atoms or molecules \cite{braaten2006uif}. Within universal few-body physics a hallmark prediction is the Efimov effect, in which three particles that interact via a resonant short-range attractive interaction exhibit an infinite series of three-body bound states, even in the regime where the two-body interaction does not support a bound state \cite{efimov1970ela}. The first experimental evidence of Efimov trimers came from an ultracold trapped gas of atoms \cite{kraemer2006efe} by tuning the strength of the interaction via a Feshbach resonance \cite{chin2010fri}. In the context of ultracold atoms, the universal regime is realized when the $s$-wave scattering length $a$, characterizing the two-body interaction in the zero-energy limit, is much larger than the characteristic range of the interaction potential. Signatures of Efimov states are imprinted on trap loss caused by three-body recombination, which typically determines the lifetime of an ultracold trapped atomic gas or Bose-Einstein condensate. So far, observations of Efimov features are observed in ultracold quantum gases of bosons: $^7$Li \cite{gross2009oou,pollack2009uit,gross2010nsi}, $^{39}$K \cite{zaccanti2009ooa}, $^{85}$Rb \cite{wild2012mot}, Cs \cite{kraemer2006efe,knoop2009ooa,berninger2011uot}, a three-spin component mixture of fermionic $^6$Li \cite{ottenstein2008cso,williams2009efa,huckans2009tbr}, and the Bose-Bose mixture $^{41}$K+$^{87}$Rb \cite{barontini2009ooh}.

In addition to the scattering length, a three-body parameter (3BP) is needed to fully describe the spectrum of Efimov trimers. The 3BP accounts for all the short-range information that is not contained in the scattering length, including a true three-body interaction. It can be parameterized as the location of the first Efimov resonance, $a_-$, on the $a$$<$0 side of a Feshbach resonance. Initially, the 3BP was thought to be very sensitive to details of the short-range interaction and therefore different for each (atomic) system \cite{dincao2009tsr}. However, experiments around different Feshbach resonances and with different alkali atoms found very similar values of the ratio $|a_-|/r_{\rm vdW}$ \cite{gross2009oou,berninger2011uot,wild2012mot}, where $r_{\rm vdW}=\frac{1}{2}(m C_6/\hbar^2)^{1/4}$ is the range of the tail of the two-body potential (also called the van der Waals length), with $m$ the atomic mass and $C_6$ the long-range coefficient. There is a vivid theoretical debate on the physical origin of this universal 3BP \cite{chin2011uso,wang2012oot,schmidt2012epb,sorensen2012epa,naidon12poo}. Most work points towards a three-body repulsive barrier that prevents the three atoms from probing the short-range interaction. An important question is how general the universal 3BP is. Refs.\,\cite{chin2011uso,wang2012oot} suggest that a two-body potential with many bound states is required, as is present in the alkali systems. However, the same 3BP was found for ground state helium-4 using a realistic two-body potential, which supports only one bound state \cite{naidon12uat}.

In this paper we investigate the possibility to extract the 3BP from our recently-measured three-body loss rate coefficient in a Bose-Einstein condensate (BEC) of metastable triplet helium-4 (denoted as \fHe) \cite{borbely2012mfd}. We will show that its value is consistent with those measured in alkali systems, providing further experimental evidence of a universal 3BP. We will also discuss other atomic systems that can be analyzed in a similar fashion. The common feature is that in the absence of a Feshbach resonance, these atomic systems already have a scattering length that is much larger than the range of the potential. The mechanism for this is an almost resonant interaction potential, i.\,e.\ a bound state is almost degenerate with the collision threshold. This potential resonance is a simple single-channel effect. In contrast, a Feshbach resonance is a multi-channel effect, where the width of the resonance introduces another length scale \cite{chin2010fri}, which may give rise to non-universal physics. Therefore, potential resonances are more directly related to the universal description connected to a large scattering length than Feshbach resonances.

\section{Three-body loss in alkalis}

To relate our work to that of the alkali experiments, we first summarize how the 3BP is extracted from three-body loss measurements around a Feshbach resonance \cite{kraemer2006efe,braaten2006uif}. In the limit of $|a|\gg r_{\rm vdW}$ the three-body loss rate coefficient $L_3$ for identical bosons is given by:
\begin{equation}
L_3 = 3 C_\pm(a)\frac{\hbar a^4}{m},
\end{equation}
where $C_\pm(a)$ are dimensionless prefactors that depend on $a$. Here we assume that three atoms are lost from the trap in the event of three-body recombination. The  scattering length $a$ is tuned by a magnetic field from $a$$>$0 to $a$$<$0 through resonance. The prefactors are given by
\begin{eqnarray}
C_+(a)=67.1e^{-2\eta_+}(\cos^2[s_0\ln(a/a_+)]+\sinh^2\eta_+)\label{Cap}\\
      +16.8(1-e^{-4\eta_+})\nonumber
\end{eqnarray}
and
\begin{equation}
C_-(a)=\frac{4590\sinh(2\eta_-)}{\sin^2[s_0\ln(a/a_-)]+\sinh^2\eta_-},\label{Cam}
\end{equation}
respectively. On top of a strong $a^4$ scaling, $L_3$ shows, as a function of $a$, a series of resonances for $a$$<$0 and minima for $a$$>$0, and the locations of these Efimov features are determined by $a_+$ and $a_-$. The parameters $\eta_\pm$ are related to the decay of the trimers into atom-dimer pairs and provide a width to the Efimov features. Experimentally $a_\pm$ and $\eta_\pm$ are obtained by fitting Eq.\,\ref{Cap} and \ref{Cam} to the measured $L_3$ spectrum as a function of $a$. For identical bosons $s_0=1.00624$, such that $C_\pm(a)=C_\pm(22.7a)$, and therefore $a_+$ and $a_-$ are defined only within a factor 22.7$^n$, $n$ being an integer.

Universal theory requires a single 3BP and therefore the Efimov features for $a$$>$0 and $a$$<$0 are related, namely via the relation $a_+/|a_-|$=0.96(3) \cite{braaten2006uif}, which has been experimentally confirmed in $^7$Li \cite{gross2009oou}. A non-universal 3BP would manifest itself as random scatter of $|a_-|$
values in a range between 1 and 22.7 for different systems. However, the ratio $|a_-|/r_{\rm vdW}$ was found in a narrow range between 8 and 10 for experiments with different alkali atoms \cite{gross2009oou,berninger2011uot,wild2012mot,wang2012oot}, indicating a universal 3BP \footnote{Note that $r_{\rm vdW}$ varies within a factor of three among the different alkali atoms.}.

\section{Analysis of three-body loss in \fHe}

Recently we have measured the three-body loss rate coefficient in a \fHe~BEC, prepared in the high-field seeking $m$=-1 Zeeman substate, and obtained the value $L_3=6.5(0.4)_{\rm stat}(0.6)_{\rm sys} \times 10^{-27}$cm$^6$s$^{-1}$ \cite{borbely2012mfd}. For spin-polarized \He~Penning ionization is strongly suppressed \cite{vassen2011cat} and three-body loss dominates the lifetime of a \fHe~BEC. Scattering of spin-polarized \He~is given by the $^5\Sigma_g^+$ potential, for which high-accuracy \emph{ab initio} electronic structure calculations are available \cite{przybytek2005bft}. For \fHe+\fHe~this potential supports 15 vibrational states. The highest excited vibrational state is weakly bound, which gives rise to a nearby potential resonance. Its binding energy is $h\times91.35(6)$\,MHz, measured by two-photon spectroscopy \cite{moal2006ado}, from which a quintet scattering length of 141.96(9)$a_0$ ($a_0$=0.05292\,nm) was deduced, consistent with the \emph{ab initio} theoretical value of 144(4)$a_0$ \cite{przybytek2005bft}. It is indeed much larger than the range of the potential, as $r_{\text{vdW}}$=35$a_0$ \footnote{Taking $C_6$=3276.680 a.\,u.\ from Ref.\,\cite{przybytek2005bft}}, such that $a/r_{\rm vdW}=4.1$. The binding energy of this weakly bound two-body state corresponds to 4.4\,mK, which is much larger than the trap depth of about 10~$\mu$K and therefore both the formed dimer and the free atom leave the trap after three-body recombination. There are no broad Feshbach resonances in \fHe~because of the absence of nuclear spin \cite{goosen2010fri}.

\begin{figure}
\includegraphics[width=8.5cm]{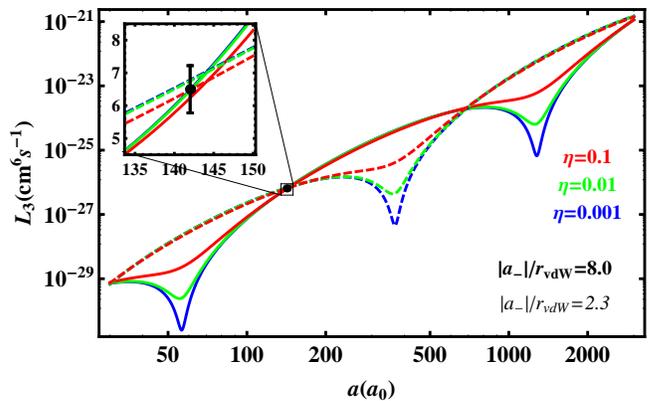}
\caption{Universal three-body loss curves (Eq.\,\ref{Cap}) for \fHe~with $|a_-|/r_{\rm vdW}$=2.3 (dashed lines) and $|a_-|/r_{\rm vdW}$=8.0 (solid lines), for different values of $\eta$, that match our measured $L_3$ value (see inset). \label{L3_vs_a}}
\end{figure}

We now consider Eq.\,\ref{Cap} to find the set of $a_+$ and $\eta_+$ values that explains our observed value of $L_3$. Following the current convention, we present the 3BP in the form $|a_-|/r_{\rm vdW}$ by using the universal relation $a_+/|a_-|$=0.96. In the alkali experiments typically $\eta_+\approx\eta_-$ and therefore in the following we will only use $\eta$. In Fig.~\ref{L3_vs_a} we show two sets of solutions of Eq.\,\ref{Cap} that match our measured $L_3$ value, namely $|a_-|/r_{\rm vdW}=2.3$ (dashed lines) and 8.0 (solid lines), for different values of $\eta$. In both cases our data point is located far outside an Efimov minimum, giving rise to a weak dependence of $\eta$ on $L_3$. That is the reason why our $L_3$ value, obtained for a single scattering length, provides information about $a_-$.

\begin{figure}
\includegraphics[width=8.5cm]{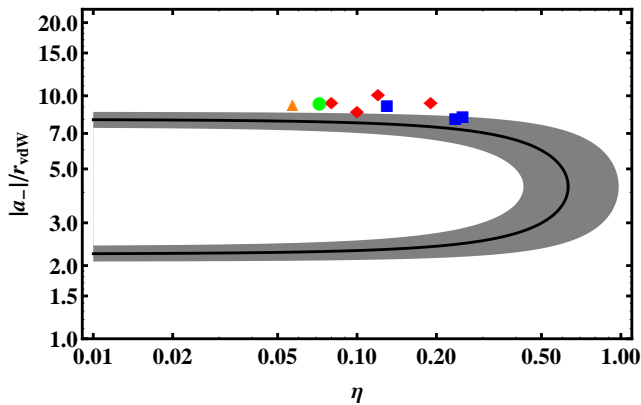}
\caption{Graphic representation of the set of $|a_-|/r_{\rm vdW}$ and $\eta$ values for which Eq.\,\ref{Cap} match our observed value of $L_3$, given by the black solid line, where the gray band corresponds to possible values based on our $L_3$ error bar. Also indicated are the obtained values for the alkali experiments: Cs \cite{berninger2011uot} (red diamond),  $^7$Li \cite{gross2009oou,pollack2009uit,gross2010nsi} (blue square), $^6$Li \cite{wenz2009uti} (green circle), $^{85}$Rb \cite{wild2012mot} (orange triangle), showing at the same time the observed range of $|a_-|/r_{\rm vdW}$ and $\eta$.\label{result_vs_eta}}
\end{figure}

In Fig.~\ref{result_vs_eta} we show the set of solutions to Eq.\,\ref{Cap} in ($|a_-|/r_{\rm vdW}$, $\eta$) parameter space for our value of $L_3$, represented by the black solid line, with the gray shaded area reflecting the experimental uncertainty in our measured $L_3$ value. Within the range of 1 to 22.7 for $|a_-|/r_{\rm vdW}$, we indeed find two narrow regions of $|a_-|/r_{\rm vdW}$ around 2.3 and 8.0, provided that $\eta$ is not too large. If $\eta$ becomes larger than 0.5 the Efimov minima are washed out and their location becomes undefined, giving rise to a broad range of possible $|a_-|/r_{\rm vdW}$ values. For comparison, the 3BP obtained from the different alkali experiments are depicted by the colored symbols. We expect the value of $\eta$ for \fHe~to be similar to those found in the alkali systems, since Penning ionization will play no important role in the decay mechanism of the Efimov trimers. Fig.~\ref{result_vs_eta} shows that our value is consistent with the 3BP found in the alkali system, considering the scatter shown in the available data and our uncertainty in $L_3$.

In our analysis we rely on two assumptions. The first assumption is that $a/r_{\rm vdW}$=4.1 is sufficiently large to apply Eq.\,\ref{Cap}. Here we notice that the three-body loss data around a Feshbach resonance fit well for $|a|$ larger than a few $r_{\rm vdW}$. Effects beyond universal theory \cite{hammer2007erc,platter2009rct,ji2010bui} may be present, but are small enough not to alter our conclusion. The second assumption is that three atoms are lost for each three-body recombination event. For $a$$>$0 additional resonances on top of the $a^4$ scaling have been observed in three-body loss spectra \cite{zaccanti2009ooa,pollack2009uit,machtey2012aoe}. Those features are explained by secondary atom-dimer collisions that are resonantly enhanced near $a=a_*$, where $a_*$ is the atom-dimer Efimov resonance position \cite{braaten2006uif}, which effectively leads to an enhancement of the number of atoms lost in a three-body recombination event. The precise underlying mechanism, and therefore what to extract from these additional resonances, is still under debate \cite{machtey2012udi,langmack2012ama,langmack2012ame}. Here we can note that if we take $|a_-|/r_{\rm vdW}$=8, then $a_*$=300$a_0$, which is far away from the actual value 142$a_0$, such that secondary atom-dimer collisions are expected not to play a role for \fHe.

\section{Other systems}

There are more atomic systems with a nearby potential resonance, for which a similar analysis as performed for \fHe~can be done once a precise measurement of $L_3$ becomes available. Alkali-metal atoms prepared in a spin-stretched state (i.\,e.\,electron and nuclear spin maximally aligned) scatter only in the triplet potential. Therefore alkalis with a large triplet scattering length provide the opportunity to extract the 3BP obtained from three-body loss in the presence of a potential resonance. Two candidates are $^{85}$Rb ($a_T$=$-$388(3)$a_0$ \cite{kempen2002ido}, $r_{\rm vdW}$=82$a_0$) and Cs ($a_T$=2440(24)$a_0$ \cite{chin2004pfs}, $r_{\rm vdW}$=101$a_0$). An experimental challenge is to distinguish three-body loss from two-body loss processes, such as spin-relaxation and hyperfine changing collisions, especially in the case of Cs \cite{soding1998gsp}.

Another group of atoms that do not possess Feshbach resonances are the alkaline-earth-metal elements and Yb. In the electronic ground state the atoms have zero electron spin and therefore there is only a single two-body potential, which is of singlet character. Furthermore, the bosonic isotopes have zero nuclear spin and two-body loss processes are completely absent. An interesting example is Ca, for which potential resonances show up for all the bosonic isotopes \cite{dammalapati2011slo}. In the following we will discuss two isotopes of Sr and Yb, for which $a$ is accurately known, $a\gg r_{\rm vdW}$ and first three-body loss measurements in BEC's have already been reported.

$^{86}$Sr ($a$=798(12)$a_0$ \cite{stein2010t1s}, $r_{\rm vdW}$=75$a_0$): Stellmer \etal~\cite{stellmer2010bec} report an upper limit of $L_3=6(3)\times10^{-24}$cm$^6$s$^{-1}$, which is one order of magnitude larger than maximally allowed by Eq.\,\ref{Cap}. The authors indicate that secondary collisions, possibly enhanced by a resonance in the atom-dimer cross section, may explain this discrepancy. We note that if one tentatively assumes that the scattering length is indeed near the atom-dimer resonance, i.\,e.\, $a_*\approx800a_0$, then $a_-\approx-750a_0$ and thus $|a_-|/r_{\rm vdW}\approx10$. This is a hint that three-body loss in $^{86}$Sr is consistent with the universal 3BP.

$^{168}$Yb ($a$=252(3)$a_0$ \cite{kitagawa2008tcp}, $r_{\rm vdW}$=78$a_0$): Sugawa \etal~\cite{sugawa2011bec} report an upper limit of $L_3=8.6(1.5)\times10^{-28}$cm$^6$s$^{-1}$. If we perform a similar analysis as for \fHe~we find again two solutions of $|a_-|/r_{\rm vdW}$. Taking the upper limit, one of the two solutions lies in a narrow range between 8 and 9. Here a smaller $L_3$ leads to a larger $|a_-|/r_{\rm vdW}$, and a value between 10 and 11 is reached when reducing the reported $L_3$ value by a factor of 2. This is a strong indication that three-body loss in $^{168}$Yb is also consistent with the universal 3BP.

\section{Conclusions}

We find our measured $L_3$ coefficient in spin-polarized \fHe~to be consistent with the 3BP that was recently found in comparing measurements using alkali-metal atoms. We give further examples of atomic systems without a Feshbach resonance but in the presence of a nearby potential resonance for which the 3BP can be extracted from an accurately-measured $L_3$, such as alkali-metal atoms in spin-stretched states and alkaline-earth atoms. We find that the three-body loss measured in $^{168}$Yb strongly indicates consistency with the universal 3BP.

We provide new experimental evidence for a universal 3BP, the first outside of the alkali-metal group and in absence of a Feshbach resonance. A universal 3BP means that short-range three-body physics is not relevant for the Efimov spectrum. This not only implies that three-body observables in the universal regime are fully determined by two-body physics, but four-body \cite{hammer2007upo,stecher2009sou,schmidt2010rgs} and $N$-body ($N$$>$4) \cite{stecher2010wbc,stecher2011fas} observables as well.

\begin{acknowledgments}
This work was financially supported by the Dutch Foundation for Fundamental Research on Matter (FOM). S.\,K.\ acknowledges financial support from the Netherlands Organization for Scientific Research (NWO) via a VIDI grant.
\end{acknowledgments}

\end{document}